\documentclass[]{spie}  

\usepackage{amsmath,amsfonts,amssymb}
\usepackage{subcaption}
\usepackage{graphicx}
\usepackage{multirow}
\usepackage[section]{placeins}
\usepackage{siunitx}
\usepackage[T1]{fontenc}
\usepackage{hhline}
\usepackage{soul}
\usepackage[colorlinks=true, allcolors=blue]{hyperref}

\usepackage[acronym]{glossaries-extra}

\setabbreviationstyle[acronym]{long-short}

\glssetcategoryattribute{acronym}{nohyperfirst}{true}

\newacronym{AU}{AU}{astronomical Unit [1.5e11 m]}  
\newacronym{pc}{pc}{parsec}
\newacronym{mas}{mas}{milliarcsecond}
\newacronym{nm}{nm}{nanometer}
\newacronym{CTE}{CTE}{coefficient of thermal expansion}
\newacronym{sqarc}{$as^2$}{square arcsecond}

\newacronym{smc}{SMC}{Small Magellanic Cloud}
\newacronym{lmc}{LMC}{Large Magellanic Cloud}
\newacronym{ism}{ISM}{interstellar medium}
\newacronym{mw}{MW}{Milky Way}
\newacronym{epseri}{$\epsilon$ Eri}{Epsilon Eridani}
\newacronym{EKB}{EKB}{Edgeworth-Kuiper Belt}
\newacronym{sed}{sed}{spectral-energy distribution}
\newacronym{CFR}{CFR}{Complete Frequency Redistribution}

\newacronym{nasa}{NASA}{National Aeronautics and Space Agency}
\newacronym{esa}{ESA}{European Space Agency}
\newacronym{omi}{OMI}{\textit{Optical Mechanics Inc.}}
\newacronym{gsfc}{GSFC}{\gls{nasa} Goddard Space Flight Center}
\newacronym{stsci}{STScI}{Space Telescope Science Institute}
\newacronym{nsroc}{NSROC}{\gls{nasa} Sounding Rocket Operations Contract}
\newacronym{wff}{WFF}{\gls{nasa} Wallops Flight Facility}
\newacronym{wsmr}{WSMR}{White Sands Missile Range}
\newacronym{BCT}{BCT}{Blue Canyon Technologies}
\newacronym{SFL}{SFL}{Spaceflight Laboratory}
\newacronym{UA}{UA}{University of Arizona}
\newacronym{UTIAS}{UTIAS}{University of Toronto Institute for Aerospace Studies}
\newacronym{DIATF}{DIATF}{Drake Imager Assembly and Testing Facility}

\newacronym{irac}{IRAC}{Infrared Array Camera}
\newacronym[plural=CCDs, firstplural=charge-coupled devices (CCDs)]{ccd}{CCD}{charge-coupled device}
\newacronym[plural=EMCCDs, firstplural=electron multiplying charge-coupled devices (EMCCDs)]{EMCCD}{EMCCD}{electron multiplying charge-coupled device}
\newacronym{DM}{DM}{Deformable Mirror}
\newacronym{MCP}{MCP}{ Microchannel Plate }
\newacronym{ipc}{IPC}{Image Proportional Counter}
\newacronym{cots}{COTS}{Commercial Off-The-Shelf}
\newacronym{COTS}{COTS}{commercial off-the-shelf}
\newacronym{ISR}{ISR}{incoherent scatter radar}
\newacronym{atcamera}{AT}{angle tracker}
\newacronym{MEMS}{MEMS}{microelectromechanical systems}
\newacronym{QE}{QE}{quantum efficiency}
\newacronym{RTD}{RTD}{Resistance Temperature Detector}
\newacronym{PID}{PID}{Proportional-Integral-Derivative}
\newacronym{PRNU}{PRNU}{photo response non-uniformity}
\newacronym{DSNU}{PRNU}{dark signal non-uniformity}
\newacronym{CMOS}{CMOS}{complementary metal–oxide–semiconductor}
\newacronym{TEC}{TEC}{thermoelectric cooler}

\newacronym{FOV}{FOV}{field-of-view}
\newacronym{NIR}{NIR}{near-infrared}
\newacronym{PV}{PV}{Peak-to-Valley}
\newacronym{MRF}{MRF}{Magnetorheological finishing}
\newacronym{AO}{AO}{Adaptive Optics}
\newacronym{TTP}{TTP}{tip, tilt, and piston}
\newacronym{FPS}{FPS}{fine pointing system}
\newacronym{SHWFS}{SHWFS}{Shack-Hartmann Wavefront Sensor}
\newacronym{OAP}{OAP}{off-axis parabola}
\newacronym{LGS}{LGS}{laser guide star}
\newacronym{WFCS}{WFCS}{wavefront control system}
\newacronym{OPD}{OPD}{optical path difference}
\newacronym{MEL}{MEL}{Master Equipment List}
\newacronym{LEO}{LEO}{low-earth orbit}
\newacronym{GEO}{GEO}{geosynchronous orbit}
\newacronym{EFC}{EFC}{electric-field conjugation}
\newacronym{iEFC}{iEFC}{implicit electric-field conjugation}
\newacronym{LDFC}{LDFC}{linear dark field control}
\newacronym{DAC}{DAC}{digital-to-analog converter}
\newacronym{TMA}{TMA}{three-mirror anastigmat}
\newacronym{ESPA}{ESPA}{EELV Secondary Payload Adapter}
\newacronym{EEID}{EEID}{Earth-equivalent Insolation Distance, the distance from the star where the incident energy density is that of the Earth received from the Sun}
\newacronym{resel}{resel}{resolution element}
\newacronym{FDPR}{FDPR}{focus diversity phase retrieval}
\newacronym{RM}{RM}{response matrix}

\newacronym{acs}{ACS}{attitude control system}
\newacronym{orsa}{ORSA}{Ogive Recovery System Assembly}
\newacronym{gse}{GSE}{ground station equipment}
\newacronym{FSM}{FSM}{fast steering Mirror}
\newacronym{CFRP}{CRFP}{carbon fiber reinforced plastic}
\newacronym{CDP}{CDP}{critical design phase}
\newacronym{CDR}{CDR}{critical design review}
\newacronym{RDI}{RDI}{reference-differential imaging}
\newacronym{ADI}{ADI}{angular-differential imaging}
\newacronym{FEA}{FEA}{finite element analysis}
\newacronym{LLOWFS}{LLOWFS}{Lyot low-order wavefront sensor}
\newacronym{WFSC}{WFS\&C}{wavefront sensing and control}
\newacronym{STOP}{STOP}{Structural-Thermal-Optical-Performance}
\newacronym{STM}{STM}{science traceability matrix}
\newacronym{ConOps}{ConOps}{concept of operations}
\newacronym{NRE}{NRE}{non-recurring engineering}
\newacronym{CSR}{CSR}{concept study report}
\newacronym{ir}{ir}{infrared}
\newacronym{XAO}{XAO}{extreme-adaptive optics}
\newacronym{AT}{AT}{angle tracking camera}
\newacronym{SRR}{SRR}{system requirements review}
\newacronym{ROI}{ROI}{region of interest}
\newacronym{LCP}{LCP}{liquid-crystal polymer}
\newacronym{CBE}{CBE}{current best estimate}
\newacronym{SBC}{SBC}{single-board computer}
\newacronym{ADCS}{ADCS}{attitude determination and control system}
\newacronym{CDH}{C$\&$DH}{command and data handling}
\newacronym{EPS}{EPS}{electrical power system}
\newacronym{TLE}{TLE}{Two Line Element set}
\newacronym{TRL}{TRL}{technology readiness level}
\newacronym{swap}{SWaP}{Size, Weight, and Power}

\newacronym{WFS}{WFS}{wavefront sensor}
\newacronym{LSI}{LSI}{Lateral Shearing Interferometer}
\newacronym{VVC}{VVC}{Vector Vortex Coronagraph}
\newacronym{VNC}{VNC}{Visible Nulling Coronagraph}
\newacronym{CGI}{CGI}{Coronagraph Instrument}
\newacronym{IWA}{IWA}{Inner Working Angle}
\newacronym{OWA}{OWA}{Outer Working Angle}
\newacronym{NPZT}{N-PZT}{Nuller piezoelectric transducer}
\newacronym{ZWFS}{ZWFS}{Zernike wavefront sensor}
\newacronym{SPC}{SPC}{Shaped Pupil Coronagraph}
\newacronym{HLC}{HLC}{Hybrid-Lyot Coronagraph}
\newacronym{LOWFSC}{LOWFS/C}{low-order wavefront sensing and control}
\newacronym{HOWFSC}{HOWFS/C}{high-order wavefront sensing and control}
\newacronym{CDI}{CDI}{coherent differential imaging}
\newacronym{SCC}{SCC}{self-coherent camera}
\newacronym{VSG}{VSG}{vacuum surface gauge}
\newacronym{TDEM}{TDEM}{Technology Development for Exoplanet Missions}
\newacronym{HZ}{HZ}{habitable zone}
\newacronym{ODI}{ODI}{orbital-differential imaging}
\newacronym{PSFTFC}{PSFTFC}{PSF template subtracted coronagraphy}
\newacronym{scoob}{SCoOB}{Space Coronagraph Optical Bench}
\newacronym{FPM}{FPM}{focal plane mask}
\newacronym{QWP}{QWP}{quarter wave plate}
\newacronym{DH}{DH}{dark hole}
\newacronym{NI}{NI}{normalized intensity}
\newacronym{DST}{DST}{Decadal Survey Testbed}

\newacronym{HST}{HST}{Hubble Space Telescope}
\newacronym{GPS}{GPS}{Global Positioning System}
\newacronym{ISS}{ISS}{International Space Station}
\newacronym[description=Advanced CCD Imaging Spectrometer]{acis}{ACIS}{Advanced \gls{ccd} Imaging Spectrometer}
\newacronym{stis}{STIS}{\textit{Space Telescope Imaging Spectrograph}}
\newacronym{mcp}{MCP}{Microchannel Plate}
\newacronym{jwst}{JWST}{$\textit{JWST}$}
\newacronym{fuse}{FUSE}{$\textit{FUSE}$}
\newacronym{galex}{GALEX}{$\textit{Galaxy Evolution Explorer}$}
\newacronym{spitzer}{Spitzer}{$\textit{Spitzer Space Telescope}$}
\newacronym{mips}{MIPS}{Multiband Imaging Photometer for \gls{spitzer}}
\newacronym{gissmo}{GISSMO}{Gas Ionization Solar Spectral Monitor}
\newacronym{iue}{IUE}{International Ultraviolet Explorer}
\newacronym{spinr}{SPINR}{$\textit{Spectrograph for Photometric Imaging with Numeric Reconstruction}$}
\newacronym{imager}{IMAGER}{$\textit{Interstellar Medium Absorption Gradient Experiment Rocket}$}
\newacronym{TPF-C}{TPF-C}{Terrestrial Planet Finder Coronagraph}
\newacronym{RAIDS}{RAIDS}{Atmospheric and Ionospheric Detection System }
\newacronym{mama}{MAMA}{Multi-Anode Microchannel Array}
\newacronym{ATLAST}{ATLAST}{Advanced Technology Large Aperture Space Telescope}
\newacronym{PICTURE}{PICTURE}{Planet Imaging Concept Testbed Using a Rocket Experiment}
\newacronym{LITES}{LITES}{Limb-imaging Ionospheric and Thermospheric
Extreme-ultraviolet Spectrograph}
\newacronym{LBT}{LBT}{Large Binocular Telescope}
\newacronym{LBTI}{LBTI}{Large Binocular Telescope Interferometer}
\newacronym{KIN}{KIN}{Keck Interferometer Nuller}
\newacronym{SHARPI}{SHARPI}{Solar High-Angular Resolution Photometric Imager}
\newacronym{IRAS}{IRAS}{Infrared Astronomical Satellite}
\newacronym{HARPS}{HARPS}{High Accuracy Radial velocity Planetary}
\newacronym{hstSTIS}{STIS}{Space Telescope Imaging Spectrograph}
\newacronym{spitzerIRAC}{IRAC}{Infrared Array Camera}
\newacronym{spitzerMIPS}{MIPS}{Multiband Imaging Photometer for Spitzer}
\newacronym{spitzerIRS}{IRS}{Infrared Spectrograph}
\newacronym{CHARA}{CHARA}{Center for High Angular Resolution Astronomy}
\newacronym{wfirst-afta}{WFIRST-AFTA}{Wide-Field InfrarRed Survey
Telescope-Astrophysics Focused Telescope Assets}
\newacronym{GPI}{GPI}{Gemini Planet Imager}
\newacronym{WFIRST}{Roman}{Nancy Grace Roman Space Telescope}
\newacronym{HabEx}{HabEx}{Habitable Exoplanet Observatory Mission Concept}
\newacronym{LUVOIR}{LUVOIR}{Large UV/Optical/Infrared Surveyor}
\newacronym{FGS}{FGS}{Fine Guidance Sensor}
\newacronym{STIS}{STIS}{Space Telescope Imaging Spectrograph}
\newacronym{MGHPCC}{MGHPCC}{Massachusetts Green High Performance
Computing Center}
\newacronym{WISE}{WISE}{Wide-field Infrared Survey Explorer}
\newacronym{ALMA}{ALMA}{Atacama Large Millimeter Array}
\newacronym{GRAIL}{GRAIL}{Gravity Recovery and Interior Laboratory}
\newacronym{jwstNIRCam}{NIRCam}{near-\gls{IR}-camera}
\newacronym{jwstMIRI}{MIRI}{Mid-Infrared Instrument}
\newacronym{Roman}{Roman}{Nancy Grace Roman Space Telescope}

\newacronym{AURIC}{AURIC}{The Atmospheric Ultraviolet Radiance Integrated Code} 
\newacronym{FFT}{FFT}{Fast Fourier Transform  }
\newacronym{MODTRAN}{MODTRAN   }{ MODerate resolution atmospheric TRANsmission }
\newacronym{idl}{IDL}{$\textit {Interactive Data Language}$}
\newacronym[sort=NED,description=NASA/IPAC Extragalactic Database]{ned}{NED}{\gls{nasa}/\gls{ipac} Extragalactic Database}
\newacronym{iraf}{IRAF}{Image Reduction and Analysis Facility}
\newacronym{wcs}{WCS}{World Coordinate System}
\newacronym{pegase}{PEGASE}{$\textit{Projet d'Etude des GAlaxies par Synthese Evolutive}$}
\newacronym{dirty}{DIRTY}{$\textit{DustI Radiative Transfer, Yeah!}$}
\newacronym{CUDA}{CUDA}{Compute Unified Device Architecture}
\newacronym{KLIP}{KLIP}{Karhunen-Lo`eve Image Processing}
\newacronym{FEM}{FEM}{finite element method}
\newacronym{CLI}{CLI}{command-line interface}
\newacronym{CAD}{CAD}{computer-aided design}
\newacronym{DBMS}{DBMS}{database management system}
\newacronym{POPPY}{POPPY}{Physical Optics Propagation in Python}

\newacronym{MSIS}{MSIS}{Mass Spectrometer Incoherent Scatter Radar}
\newacronym{nmf2}{$N_m$}{F2-Region Peak density}
\newacronym{hmf2}{$h_m$}{F2-Region Peak height}
\newacronym{H}{$H$}{F2-Region Scale Height}

\newacronym{isr}{ISR}{Incoherent Scatter Radar}
\newacronym[description=TLA Within Another Acronym]{twaa}{TWAA}{\gls{tla} Within Another Acronym}
\newacronym[plural=SNe, firstplural=Supernovae (SNe)]{sn}{SN}{Supernova}
\newacronym{EUV}{EUV}{Extreme-Ultraviolet }
\newacronym{EUVS}{EUVS}{\gls{EUV} Spectrograph}
\newacronym{F2}{F2}{Ionospheric Chapman F Layer }
\newacronym{F10.7}{F10.7}{ 10.7 cm radio flux [10$^{-22}$ W m$^{-2}$ Hz$^{-1}$]  }
\newacronym{FUV}{FUV}{far-ultraviolet }
\newacronym{IR}{IR}{infrared}
\newacronym{MUV}{MUV}{mid-ultraviolet }
\newacronym{NUV}{NUV}{near-ultraviolet }
\newacronym{nir}{NIR}{near-infrared}
\newacronym{mir}{MIR}{mid-infrared}
\newacronym{uv}{UV}{ultraviolet}
\newacronym{O$^+$}{O$^+$}{Singly Ionized Oxygen  Atom }
\newacronym{OI}{OI}{Neutral Atomic Oxygen Spectroscopic State }
\newacronym{OII}{OII}{Singly Ionized Atomic Oxygen Spectroscopic State }
\newacronym{PSF}{PSF}{point spread function}
\newacronym{$R_E$}{$R_E$}{Earth radii [$\approx$ 6400 km]  }
\newacronym{RV}{RV}{radial velocity}
\newacronym{UV}{UV}{ultraviolet }
\newacronym{WFE}{WFE}{wavefront error}
\newacronym[plural=PAHs, firstplural=Polycyclic Aromatic Hydrocarbons (PAHs)]{pah}{PAH}{Polycyclic Aromatic Hydrocarbon}
\newacronym{obsid}{OBSID}{Observation Identification}
\newacronym{SZA}{SZA}{Solar Zenith Angle}
\newacronym{DOF}{DOF}{degrees-of-freedom}
\newacronym{PZT}{PZT}{lead zirconate titanate}

\newacronym{PCA}{PCA}{principal component analysis}
\newacronym{fwhm}{FWHM}{full-width-half maximum}
\newacronym{RMS}{RMS}{root mean squared}
\newacronym{RMSE}{RMSE}{root mean squared error}
\newacronym{MCMC}{MCMC}{Marcov chain Monte Carlo}
\newacronym{DIT}{DIT}{discrete inverse theory}
\newacronym{SNR}{SNR}{signal-to-noise ratio}
\newacronym{PSD}{PSD}{power spectral density}
\newacronym{NMF}{NMF}{non-negative matrix factorization}
\newacronym{SCoOB}{scoob}{space-coronagraph optical bench}

\title{The space coronagraph optical bench (SCoOB): 4. vacuum performance of a high contrast imaging testbed}
\authorinfo{Send correspondence to kvangorkom@arizona.edu}

\author[a]{Kyle Van Gorkom}
\author[a]{Ewan S Douglas}
\author[a,b]{Kian Milani}
\author[a,b]{Jaren N Ashcraft}
\author[a]{Ramya M Anche}
\author[a,b]{Emory Jenkins}
\author[a]{Patrick Ingraham}
\author[a]{Sebastiaan Haffert}
\author[b]{Daewook Kim}
\author[b]{Heejoo Choi}
\author[a]{Olivier Durney}
\affil[a]{Steward Observatory, University of Arizona, 933N Cherry Avenue, Tucson, Arizona, 85721, USA}
\affil[b]{James C. Wyant College of Optical Sciences, University of Arizona, 933N Cherry Avenue, Tucson, Arizona, 85721, USA}

\begin{document} 

\maketitle

\begin{abstract}

The Space Coronagraph Optical Bench (SCoOB) is a high-contrast imaging testbed built to demonstrate starlight suppression techniques at visible wavelengths in a space-like vacuum environment. The testbed is designed to achieve ${<}10^{-8}$ contrast from $3-10\lambda/D$ in a one-sided dark hole using a liquid crystal vector vortex waveplate and a 952-actuator Kilo-C deformable mirror (DM) from Boston Micromachines (BMC). We have recently expanded the testbed to include a field stop for mitigation of stray/scattered light, a precision-fabricated pinhole in the source simulator, a Minus K passive vibration isolation table for jitter reduction, and a low-noise vacuum-compatible CMOS sensor. We report the latest contrast performance achieved using implicit electric field conjugation (iEFC) at a vacuum of ${\sim}10^{-6}$ Torr and over a range of bandpasses with central wavelengths from 500 to 650nm and bandwidths (BW) from $\ll 1\%$ to 15\%. Our jitter in vacuum is $<3\times10^{-3} \lambda/D$, and the best contrast performance to-date in a half-sided D-shaped dark hole is $2.2\times10^{-9}$ in a $\ll 1 \%$ BW, $4\times10^{-9}$ in a 2\% BW, and $2.5\times10^{-8}$ in a 15\% BW.

\end{abstract}


\section{INTRODUCTION}
\label{sec:intro}

Space-based astronomical telescopes, lacking atmospheric seeing effects, have long promised the image stability needed to resolve planets around other stars (e.g. \cite{kenknight_methods_1977, breckinridge_space_1984}). Unfortunately, coronagraph technology to remove host star light and resolve planets has taken many decades to develop in the lab. Trauger and Traub\cite{trauger_laboratory_2007-1} first demonstrated $<10^{-9}$ contrast in laser light in 2007 using a Lyot coronagraph, and the Roman Coronagraph mission expected to launch in 2027 will demonstrate active wavefront control and planet-star flux ratio sensitivity better than $10^{-8}$ on a space observatory \cite{bailey_nancy_2023}. 
Much work remains to reach Earthlike exoplanet flux ratio sensitivities below $10^{-10}$. 
This includes demonstration of coronagraphs that are high-throughput, effective across a large bandwidth, robust to telescope errors, and operationally efficient in space. 

A primary goal of \gls{scoob} is to raise the \gls{TRL} of integrated coronagraph hardware and software systems including pointing control, wavefront control, coronagraph masks, low-order and high-order wavefront sensing, sensors and compute platforms. We seek to combine components that have either been demonstrated in the lab or in space into a functional instrument operating in a space-like environment.
We seek to mature existing technologies by combining flight ready subsystems, with a focus on exploring physical limitations independent of telescope aperture.
We are also collaborating with ground-based \gls{AO} groups to develop flexible Linux-based coronagraph control software to enable rapid deployment of future instruments. 
The design of SCoOB was introduced in Ashcraft et al.\cite{ashcraft_space_2022} and first light was presented in Van Gorkom et al\cite{vangorkom_scoob_2022}.
Here we present an overview of recent activities on scoob, including updates to the optical layout, the integration of new \glspl{FPM}, and contrast performance in air and in vacuum.


\section{TESTBED STATUS}\label{sec:status}

\begin{figure}
    \centering
    \includegraphics[width=\textwidth]{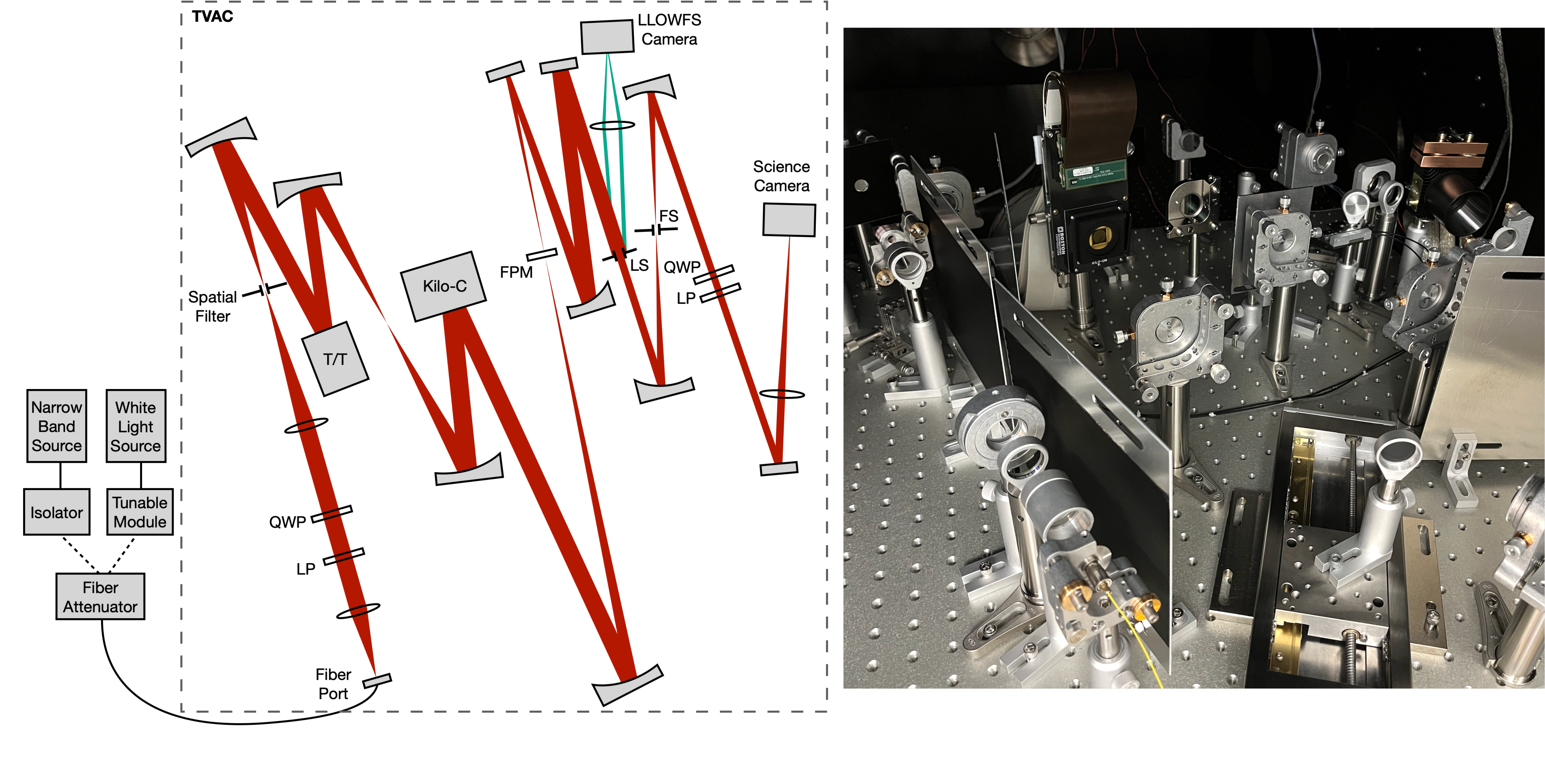}
    \caption{Left: Optical layout of \glsxtrshort{scoob}, as of June 2024. A source is fiber-fed into the TVAC chamber, circularly polarized with a linear polarizer (LP) and \glsxtrshort{QWP}, and relayed to a spatial filter with a pair of achromatic doublets. The entrance pupil is defined at the tip/tilt (T/T) mirror and re-imaged at the Kilo-C DM. The \glsxtrshort{FPM} is placed in an f/48 beam. A 95\% Lyot stop is placed in a conjugate pupil, and the reflected beam is relayed to the \glsxtrshort{LLOWFS}. The transmitted beam passes through a field stop (FS) in a focal plane, followed by a circular analyzer (QWP and LP), and is finally focused onto the science camera. Right: Photo of SCoOB in the TVAC chamber in May 2024.}
    \label{fig:testbed_layout}
\end{figure}

\begin{figure}
    \centering
    \includegraphics[width=\textwidth]{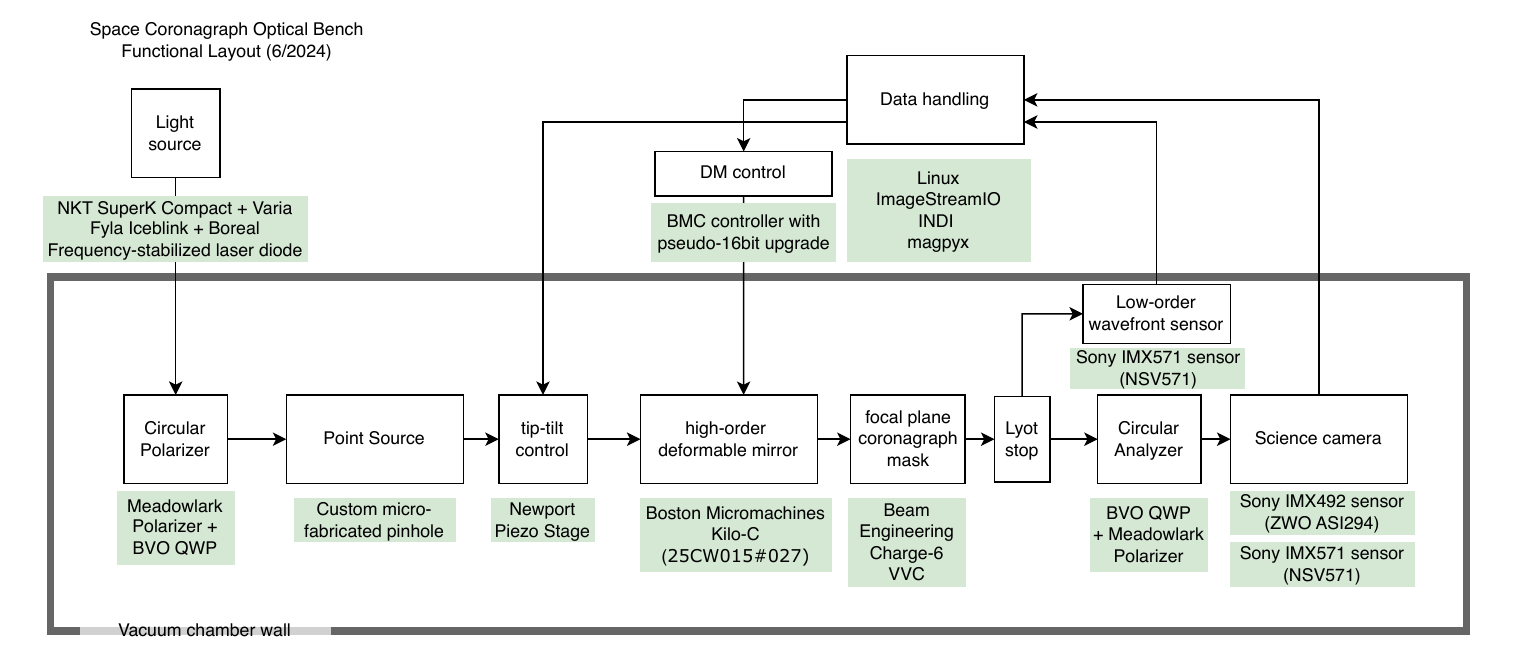}
    \caption{Functional blocks of the testbed. Typical configuration shown in green boxes beneath general functionalities.}
    \label{fig:functional_blocks}
\end{figure}

The detailed optical design of \gls{scoob} is presented in Ashcraft et al.\cite{ashcraft_space_2022}, but the layout of the testbed has been significantly expanded and modified since. Early \gls{WFSC} efforts including phase retrieval and dark-hole digging with a knife-edge coronagraph and \gls{iEFC} were presented in Van Gorkom et al.\cite{vangorkom_scoob_2022} An updated optical layout of SCoOB is shown in Figure \ref{fig:testbed_layout}, and a functional block diagram of the testbed is given in Figure \ref{fig:functional_blocks}. The key features of the testbed, along with a description of recent upgrades and current efforts, are outlined below.

In its current configuration, SCoOB features a 952-actuator Kilo-C \gls{DM} from Boston Micromachines Corp. (BMC) (with a single bad actuator stuck at the 0V position), a \gls{VVC} manufactured by Beam Engineering Company, and a low read-noise, vacuum-compatible CMOS sensor. The \glspl{OAP} that constitute the majority of the relay optics are protected aluminum (Al+MgF2) on a Zerodur substrate with high throughput into the UV and low roughness ($\lesssim 0.5$ nm RMS surface error on spatial scales smaller than 0.08mm). A multi-layered mechanical isolation system including pneumatic legs through the chamber and a Minus K platform maintains jitter $\lesssim 0.003 \lambda/D$ RMS under vacuum. It was recently expanded to accommodate a field stop for additional scattered light mitigation.

The testbed control software is built around MagAO-X\cite{magaox}, which relies on \texttt{cacao}\cite{cacao} for low-latency operations (e.g., reading cameras, commanding DMs) and on  Instrument-Neutral Distributed
Interface (INDI)\cite{indi} for high-latency operations (e.g., commanding stages, setting camera properties). Most interactive \gls{WFSC} activities on the testbed, including \gls{DH}-digging, are performed via Python interfaces with shared memory images. 

The \gls{DM} controller is the \gls{COTS} controller supplied by BMC, supplemented with a low-noise pseudo 16-bit filter board capable of a minimum actuator step size of 20 pm. Quantization error at this level is expected to limit the achievable contrast to $<2\times10^{-10}$ at 650nm \cite{ruane_2021_quantization}. The beam size at the DM is ${\sim}$9.6mm for about 32 actuators across the pupil.

At the front end, an expanded source simulator was designed to allow spatial filtering of the fiber output. A pair of achromatic lenses relay the fiber to a $4\si{\micro \meter}$-diameter micro/nano-fabricated pinhole produced in-house at UArizona\cite{jenkins_microfabricated_2023}. With a 6.8mm entrance pupil and 147mm focal length on the first \gls{OAP}, the $4\si{\micro \meter}$ pinhole is unresolved, with a diameter of $30\%\ \lambda/D$ at 630nm. The pupil stop is a flat mirror masked with an Aeroglaze Z306 coating to define the circular aperture and mounted on a vacuum compatible piezo-actuator tip/tilt stage.

The source is fed into the vacuum chamber via a single-mode fiber potted in a vacuum feedthrough. Several sources were coupled into the fiber for the work presented here. We use two supercontinuum light sources paired with wavelength tuners: the Fyla Iceblink with Boreal tuner, and the NKT SuperK Compact paired with the Varia tuner. Both sources can be tuned to wavelengths in the ${\sim}$450-750nm range with bandwidths from ${\sim}$5 to >100nm. For narrow-band experiments, we use a 632.8nm frequency-stabilized laser diode. The narrow-band source is fed through an inline fiber isolator to minimize feedback from a back-reflection from the pinhole, which was discovered to cause instability in the source. The sources are fed through a fiber attenuator with over 5 decades of power attenuation for precise control over the input power.

The optical layout downstream of the \gls{FPM} has been entirely reworked to eliminate as many transmissive optics as possible and to accommodate a field stop for the mitigation of stray and scattered light. A single achromatic doublet is used as the last optic to form an image at the science camera.

To mitigate stray light, a set of aluminum panels with Aktar light absorbent foil were placed at strategic locations. Large panels were placed alongside the source simulator optics, as well as in the direction faced by the science camera. Smaller panels with cut-outs for optics were placed immediately in front of surfaces where the beam was known to overfill the optics---the first OAP after the spatial filter, and the fold flat and OAP downstream of the FPM.

A significant difference between the previously presented results and this work is the change from a knife-edge coronagraph to a \gls{VVC}. In  lab testing elsewhere, unobscured VVCs \cite{mawet_annular_2005,foo_optical_2005,mawet_optical_2009,ruane_vortex_2018,Serabyn19} have  demonstrated a raw contrast of $1.6\times10^{-9}$ in a 10\% bandwidth \cite{ruane_broadband_2022}, and are less sensitive to low-order aberrations and stellar diameters than classical Lyot coronagraphs \cite{ruane_vortex_2018}. This combination of properties: starlight suppression, planet transmission, bandwidth, and robustness to 
aberrations have led the VVC to be baselined by HabEx \cite{gaudi_habitable_2020} and CDEEP mission concepts \cite{maier_design_2020} and were recently demonstrated at the $5 \times 10^{-6}$ level suborbitally by PICTURE-C balloon \cite{mendillo_picture-c_2023}. 

\begin{figure}[t]
    \centering
    \includegraphics[width=0.6\linewidth]{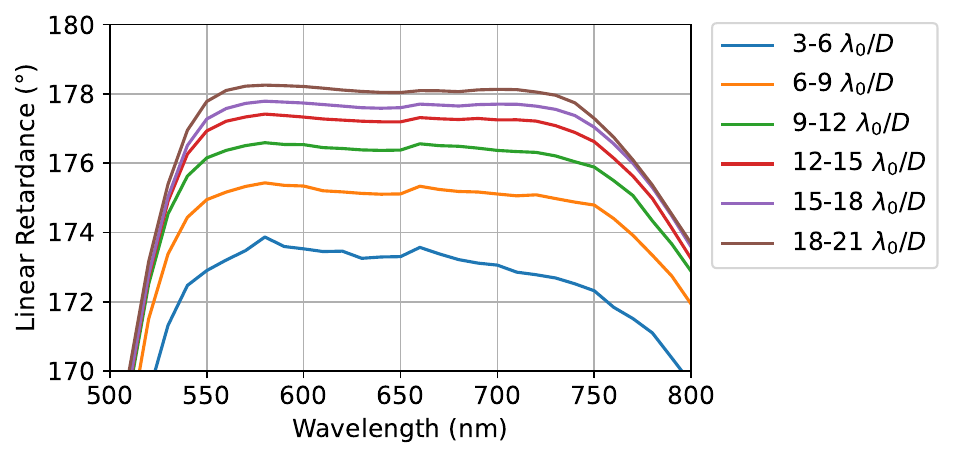}
    \caption{Mean linear retardance of the \gls{VVC} unit used in the results presented here, in binned radial separations from the center of the device. Here, $\lambda_0$ = 630nm. More detailed analysis of this and other units is given in Anche et al.\cite{anche_pol_scoob_2024} Data from Beam Engineering Company.}
    \label{fig:vvc_unit}
\end{figure}

The \gls{FPM} is a charge-6 VVC intended to operate over a 20\% bandwidth centered at 635nm manufactured by Beam Engineering Company. 
The VVCs are mounted on Ohara S-BAL35R substrates, manufactured by Rainbow Research Optics, a non-browning radiation resistant glass with an index of refraction of 1.59. Figure \ref{fig:vvc_unit} shows the linear retardance of a unit at UArizona as a function of radial separation from the singularity at the center of the masks. To filter the leakage term that arises from imperfect half-wave retardance on the \gls{VVC}, a pair of Meadowlark linear polarizers and Bolder Vision Optiks \glspl{QWP} are placed in the source optics and after the field stop to form a circular polarizer and analyzer. See Anche et al.\cite{anche_pol_scoob_2024} for a detailed description of these devices and modeling of polarization aberrations in \gls{scoob}, and Ashcraft et al.\cite{ashcraft_mueller_scoob_2024} for in-situ Mueller matrix measurements of the testbed.

Two Lyot stops are in active use in the testbed: one is a 95\% aluminum stop painted with Aeroglaze Z306 with an angled knife edge to minimize glint, and the second is a reflective stop fabricated at UMass Lowell via a lithography process to relay the beam outside the geometrical pupil to the \gls{LLOWFS} for dark hole stabilization \cite{mendillo_llowfs_2023}.

Two cameras were used for the results reported in these proceedings. In air, we use a ZWO ASI 294MM Pro camera, a Sony IMX492 CMOS chip with 2.3$\si{\micro \meter}$ pixels, $\sim$2 e$^{-}$ read noise in high conversion gain mode, and a 14-bit ADC. In vacuum, we employ a Sony IMX571---3.76$\si{\micro \meter}$ pixels, $<1.5$e$^{-}$ read noise in high conversion gain mode, and a 16-bit ADC---packaged by Neutralino Space Ventures (NSV) for vacuum operations.

To monitor thermal stability while in vacuum, we installed four temperature sensors at keys point: two at different locations on the camera housing, one on the upper bench, and one at the lower bench at the base of the sensor heat straps.

\section{Vacuum Chamber}\label{sec:tvac}

Tests were performed in the basement of Steward Observatory, on the University of Arizona main campus, in a 1.2 m diameter, 2.2 m deep stainless steel vacuum chamber manufactured by Rydberg Vacuum Sciences in Renton, Washington, USA.
The interior of the chamber is isolated from the wall by a thermal shroud, which is coated with Akzo Nobel 463-3-8 Flat Black epoxy.  
The exterior of the chamber is fiberglass insulated and covered in stainless steel sheet metal. While the chamber provides active thermal control, the chamber temperature was passively controlled during these tests. 
The chamber is designed to reach $10^{-8}$ Torr using a cryopanel. In non-cryo operation using a  turbomolecular pump, the vacuum approaches $10^{-6}$ Torr.

\begin{figure}
    \centering
    \includegraphics[width=\textwidth]{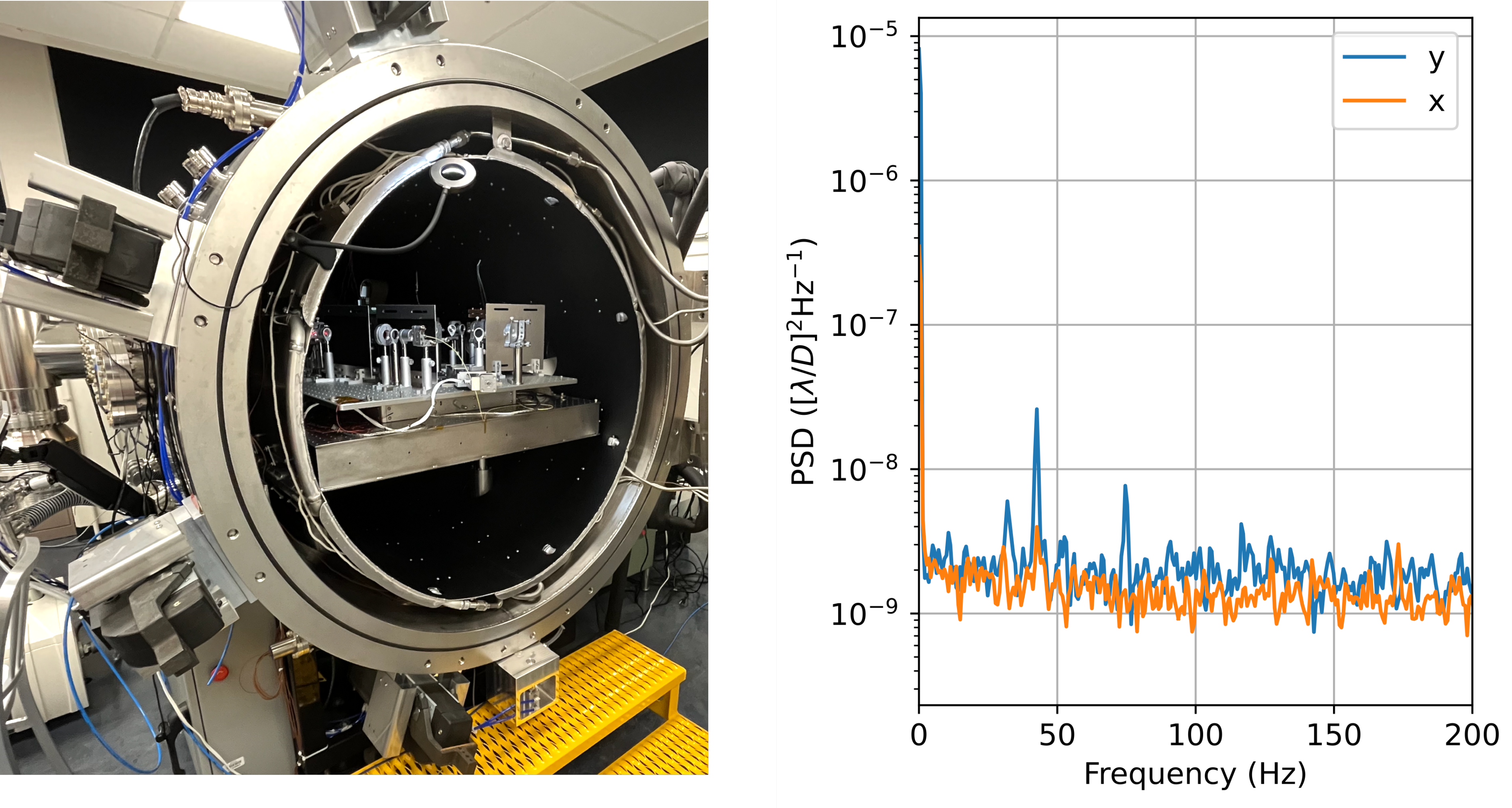}
    \caption{Left: \gls{scoob} in TVAC chamber. Right: Jitter \gls{PSD} measured on the science camera at vacuum, in horizontal (x) and vertical (y) directions. RMS jitter is $<3\times10^{-3}\lambda/D$ and is dominated by very slow drifts.}
    \label{fig:scoob_tvac}
\end{figure}

Vibration isolation in the chamber is accomplished by floating a 4 inch vacuum compatible optical breadboard on a table attached to pneumatic bellows exterior to the chamber. A final stage of vibration isolation is accomplished with a passive Minus-K CT-2 vibration isolation table. 
Measured vacuum jitter at the science camera is $<3\times10^{-3}\lambda/D$. 
The right panel of Figure \ref{fig:scoob_tvac} shows the \gls{PSD} of measured jitter in the testbed. 

\section{ALIGNMENT AND CALIBRATION}\label{sec:align}

To set the input circular polarization, we place a diffraction polarization grating---which splits the incident light into orthogonal circular polarization states in the +1 and -1 orders---in the f/48 beam just upstream of the \gls{FPM}. With a camera at the FPM position, we rotate the input \gls{QWP} until one of the diffraction orders is minimized. With the polarization grating and FPM removed from the beam, we rotate the output QWP to drive the intensity at the science camera to a minimum, thereby setting the circular analyzer to the orthogonal circular polarization.

Prior to placing the FPM in the beam, we run a closed-loop \gls{FDPR} correction\cite{thurman, van_gorkom_space_2022} with a camera at the FPM position to minimize the \gls{WFE} incident on the \gls{VVC}. Defocus modes on the \gls{DM} create the defocus diversity required to break the degeneracy in the phase solution. The FDPR response matrix is constructed from a Hadamard basis\cite{kasper_2004} measured on the DM and pseudo-inverted with a truncated Tikhonov regularization to form the control matrix. The WFE typically converges to $>95\%$ Strehl ratio within 5 iterations.

The FPM is aligned in a two-step process. In step one, following the technique outlined in Ruane et al.\cite{ruane_broadband_2022}, we place a ``diffuser'' lens upstream of the FPM to create an (approximately) uniform illumination on the FPM, which is re-imaged at the science camera. The FPM is shifted along the optical axis until diffraction pattern from the opaque spot at the center of the mask is minimized, indicating that the FPM is in a conjugate focal plane. To refine this solution, we re-image a downstream pupil on a camera and measure the summed intensity within the geometrical pupil as we command the DM over a range of defocus values. When the FPM is at focus, this signal is centered at a DM defocus amplitude of 0; when the FPM is slightly out of focus, the signal shows a strong asymmetry, and the direction of the asymmetry informs the direction the FPM needs to move.

Once the FPM is aligned, the Lyot stop is inserted and aligned using the pupil viewer such that it is centered within the bright ring formed by the VVC. The field stop is aligned by minimizing the diffraction from the edges of the stop on the science camera, and is mounted on a horizontal linear stage that allows it to be moved in and out of the beam. The source is aligned to the FPM by driving the tip/tilt mirror until the measured intensity on the science camera in the region immediately surrounding the FPM singularity is minimized. The plate scale (in $\lambda/D$) at the science focal plane is determined by measuring the locations of the DM actuator print-through speckles and accounting for the measured beam footprint on the DM. This value agrees with the plate scale expected from the optical design of the system to better than $1\%$.

\section{CONTRAST PERFORMANCE}\label{sec:contrast}

\begin{figure}
    \centering
    \includegraphics[width=1\linewidth]{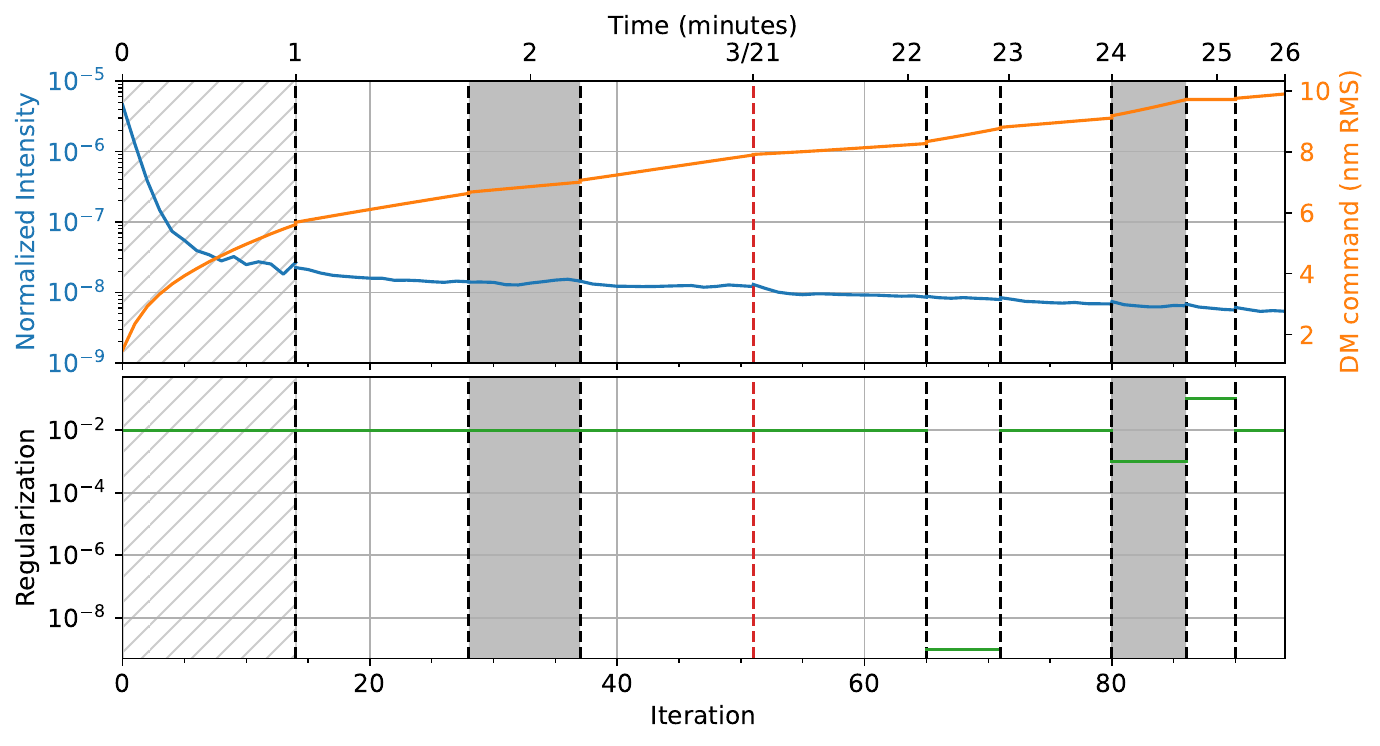}
    \caption{Normalized intensity, RMS \gls{DM} surface, and regularization term for a typical DH-digging sequence with \gls{iEFC}. Dashed lines mark iterations where control terms were changed. In the first 15 iterations (hatched region), the control region is an oversized D-shaped \gls{DH} from 2.5 to 11$\lambda/D$, and a leaky integrator with a 10\% leak is used. After the first 15 iterations, the DH region is reduced to 3 to 10$\lambda/D$ and the leak is set to 0. In the shaded regions, the uniform-weighted D-shaped DH mask is replaced with a mask weighted by the residual signal in the DH to target stubborn speckles. The red dashed line indicates where a new iEFC Jacobian was measured (analogous to a re-linearization of the Jacobian with \glsxtrshort{EFC}). The green line on the lower plot shows modulation of the Tikhonov regularization term. Time is given in minutes since the start of the DH-digging (excluding the initial iEFC Jacobian calibration). The jump in time from 3 to 21 minutes is due to the length of the second iEFC calibration step.}
    \label{fig:dh_sequence}
\end{figure}

\begin{figure}
\vspace{-4\baselineskip}
    \centering
	\includegraphics[width=0.9\linewidth]{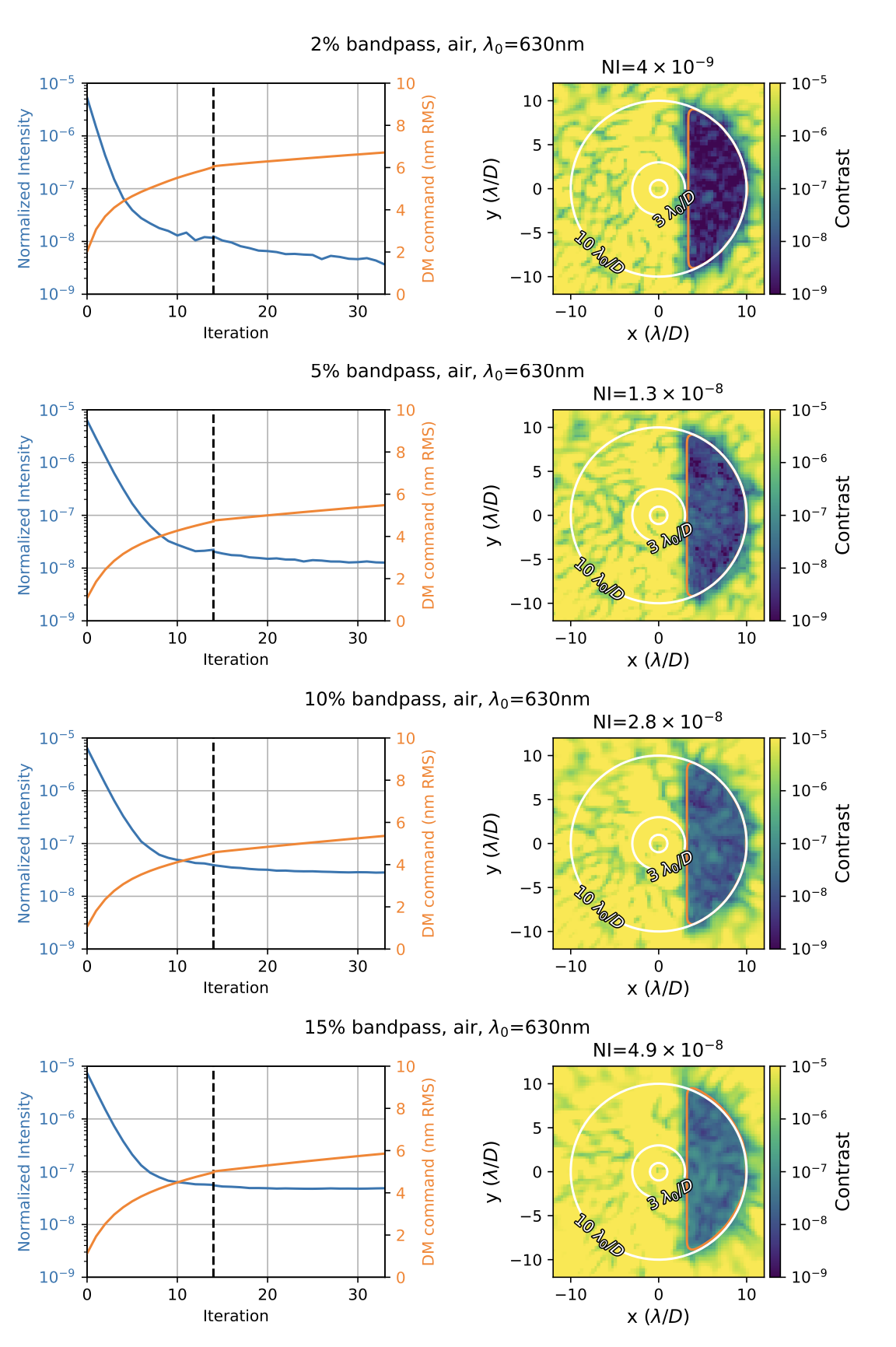}
\caption 
{ \label{fig:dh_air} Testbed performance in air at a central wavelength of $\lambda_0=630$nm in bandwidths from 2-15\%. Left column: \gls{NI} and RMS \gls{DM} surface versus iteration for each of the cases. The dashed line marks a re-calibration of the \gls{iEFC} Jacobian.  Right column: DH images and mean NI in a D-shaped \gls{DH} from 3-10$\lambda_0/D$. Note that in this case, the iEFC measurements were made in the full bandpass. Compare to Figure \ref{fig:dh_vacuum_bband}, where iEFC was performed in a narrow bandpass but evaluated in broader bandpasses.}
\end{figure} 

With the source aligned to the center of the \gls{FPM}, the typical normalized intensity between 3 and 10$\lambda/D$ prior to any dark-hole digging is approximately $5\times10^{-6}$ to $1\times10^{-5}$. To dig to a deeper contrast in a half-sided, D-shaped dark hole over 3-10$\lambda/D$, we use \gls{iEFC}\cite{iefc}, a model-free variant of \gls{EFC}\cite{efc} that uses double-difference probe images to construct an empirical Jacobian without a direct estimate of the electric field and attempts to minimize intensity within the targeted region.

iEFC requires two sets of probes: pairwise probes to modulate the focal-plane intensity and sense the term proportional to the electric field, and calibration probes to build the iEFC \gls{RM}. For the calibration probes, we adopt a truncated Hadamard basis\cite{kasper_2004}, the amplitude of which is chosen heuristically dependent on the contrast of the underlying speckle field. Typical values are 2nm for speckles at the $10^{-6}-10^{-5}$ contrast level, and 0.5nm for $10^{-9}-10^{-8}$ contrast. For the pairwise probes, we choose a set of 3 probes generated from a power law spectrum of the form $r^{\alpha}$, where $r$ is cycles/aperture and $\alpha$ is typically set to $2$ to compensate for the fall-off of the DM transfer function toward higher spatial frequencies. The pairwise probes are normalized to an amplitude similar to the calibration probes. For each sequential calibration, the probes are randomized to create some diversity in the probe speckle patterns and reduce the possibility of null- or poorly-sensed modes in the focal plane. The double-difference probe patterns are concatenated into a single vector and pseudo-inverted with a truncated Tikhonov regularization to build the iEFC control matrix.

To avoid stagnation in the digging process, we vary the regularization similar to the beta-bumping technique employed in EFC\cite{seo_beta_bumping}, as well as the focal-plane weighting. In initial iEFC iterations, we attempt to minimize the pairwise-probed signal in an oversized (e.g., 2-12$\lambda/D$), uniform focal-plane region with a leaky integrator (with a typically 10\% leak). In subsequent iterations, we shrink the weighted region to the dark zone of interest (e.g., 3-10$\lambda/D$) and set the leak to 0. Tikhonov regularization values are typically set to $10^{-2}$ but are interspersed with values ranging from $10^{-9}$ to $10^{-1}$. We find that iEFC often converges to solutions that leave coherent speckles in the dark hole, and we target these by re-building the control matrix with a dark zone focal-plane mask weighted by the measured dark hole intensity (residual speckles). An example of an iEFC digging sequence employing these techniques is shown in Figure \ref{fig:dh_sequence}.

We report contrast in terms of normalized intensity (as defined in Mennesson et al.\cite{mennesson_2024}), the ratio of the counts in the dark zone to the peak counts of the source with the coronagraph mask removed; rather than remove the \gls{VVC}, however, we measure the normalization term by driving the \gls{PSF} off-axis by ${\sim}25 \lambda/D$. In the following sections, we explore the contrast performance of our system in both air and vacuum, at wavelengths as short as 500nm, and in bandwidths up to 15\%.

\subsection{Results}

\begin{table}
\centering
\setlength\doublerulesep{0.2cm} 
\begin{tabular}{ |c|c|c|  }
 \hline
 \multicolumn{3}{|c|}{$\lambda_0$=630nm, $\ll$1 - 15\% bandwidths} \\
 \hline
 Bandwidth & Mean NI (3-10$\lambda_0/D$) & Comments\\
 \hline
 $\ll$1\%   & $2.2 \times 10^{-9}$  & WFS\&C at $\ll$1\% BW\\
 \hline
 2\%      & $4.0 \times 10^{-9}$   & WFS\&C at 2\% BW, air, no field stop\\
\hline
 5\%      & $8.2 \times 10^{-9}$   & WFS\&C at $\ll$1\% BW\\
 \hline
 10\%     & $1.5 \times 10^{-8}$   & \textquotedbl \\
 \hline
 15\%     & $2.5 \times 10^{-8}$   & \textquotedbl \\
 \hline
 \hline
 \multicolumn{3}{|c|}{2\% sub-bands over a 10\% bandwidth at $\lambda_0$=630nm} \\
 \hline
 $\lambda_\mathrm{min}$ - $\lambda_\mathrm{max}$ (nm) & Mean NI (3-10$\lambda_0/D$)  & Comments\\
 \hline
 598.5 - 611.1   & $3.1 \times 10^{-8}$      & WFS\&C at $\ll$1\% BW, $\lambda_0$=630nm\\
 \hline
611.1 - 623.7      & $1.3 \times 10^{-8}$    &  \textquotedbl \\
\hline
 623.7 - 636.3     & $4.0 \times 10^{-9}$    &  WFS\&C at 2\% BW, air, no field stop \\
 \hline
 636.3 - 648.9     & $9.8 \times 10^{-9}$    &  WFS\&C at $\ll$1\% BW, $\lambda_0$=630nm \\
 \hline
648.9 - 661.5     & $2.3 \times 10^{-8}$     &  \textquotedbl \\
 \hline
 \hline
 \multicolumn{3}{|c|}{2\% bandwidths} \\
 \hline
 $\lambda$ (nm) & Mean NI (3-10$\lambda/D$)  & Comments\\
 \hline
493   & $2.5 \times 10^{-8}$             & outside of VVC design bandpass \\
 \hline
543     & $1.3 \times 10^{-8}$           &  outside of VVC design bandpass \\
\hline
 630     & $4.0 \times 10^{-9}$          &  WFS\&C at 2\% BW, air, no field stop\\
 \hline
\end{tabular}
\caption{
\label{table:contrast_summary}
Summary of the best contrast performance to date on \gls{scoob} in a number of configurations. Note that all results in this table were captured with the testbed in vacuum, with the exception of the 2\% BW centered at 630nm, which is repeated in each category for reference.
}
\end{table}

\begin{figure}[!htb]
    \centering
    
	\includegraphics[width=0.9\linewidth]{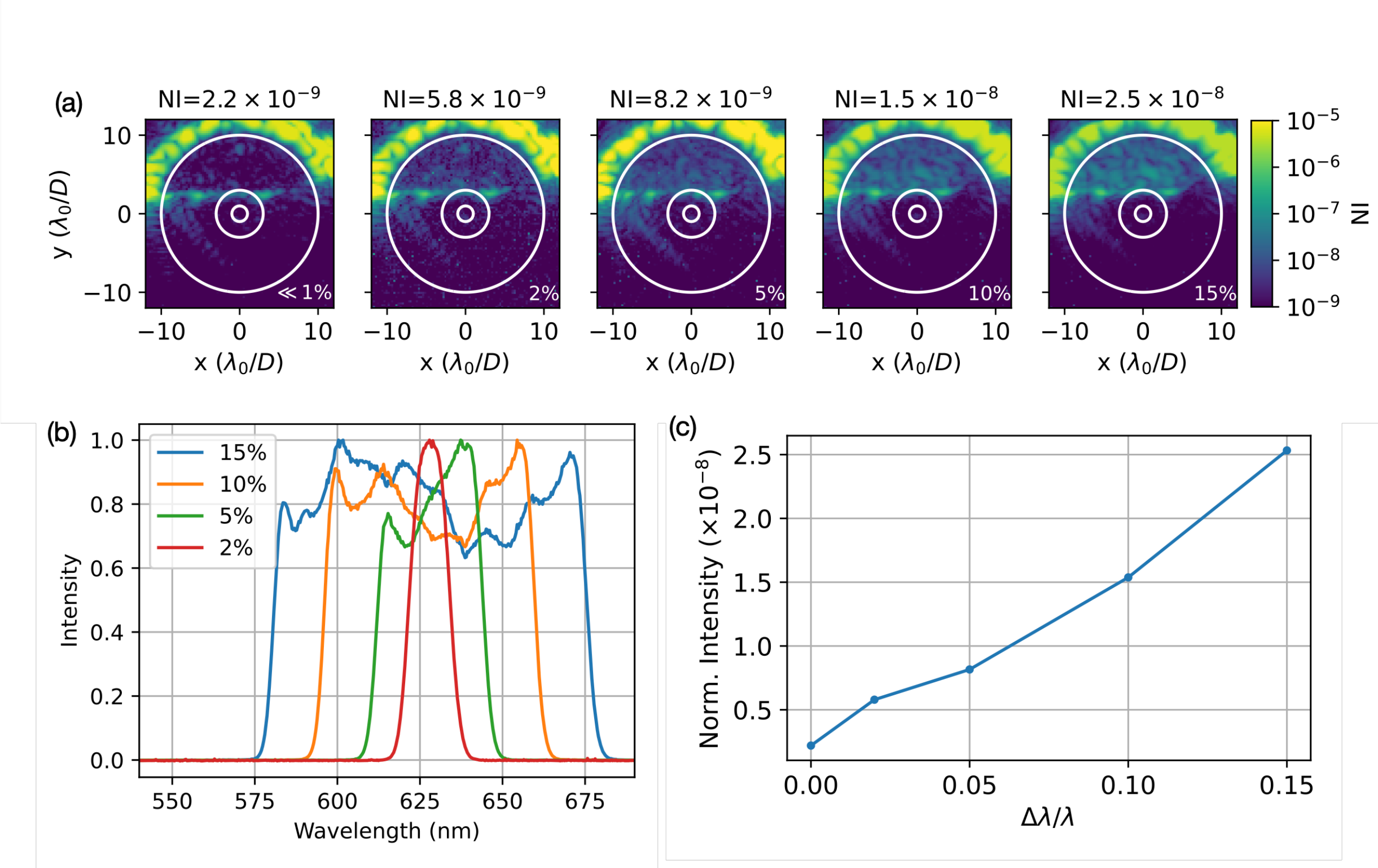}
    \caption 
    { \label{fig:dh_vacuum_bband} (a) NI images of \glspl{DH} in bandwidths from $\ll1\%$ to $15\%$, with a central wavelength of $\lambda_0=$630nm. Radial separations of 1, 3, and 10 $\lambda_0/D$ are marked. (b) Measured spectra of each of the bandpasses. Note that the spectrum corresponding to the $\ll1\%$ BW is excluded, as the measured width of the narrowband source is limited by our spectral resolution. (c) The mean \gls{NI} in a half-sided DH from $3-10\lambda_0/D$ as a function of spectral bandwidth.}
    
	\includegraphics[width=0.9\linewidth]{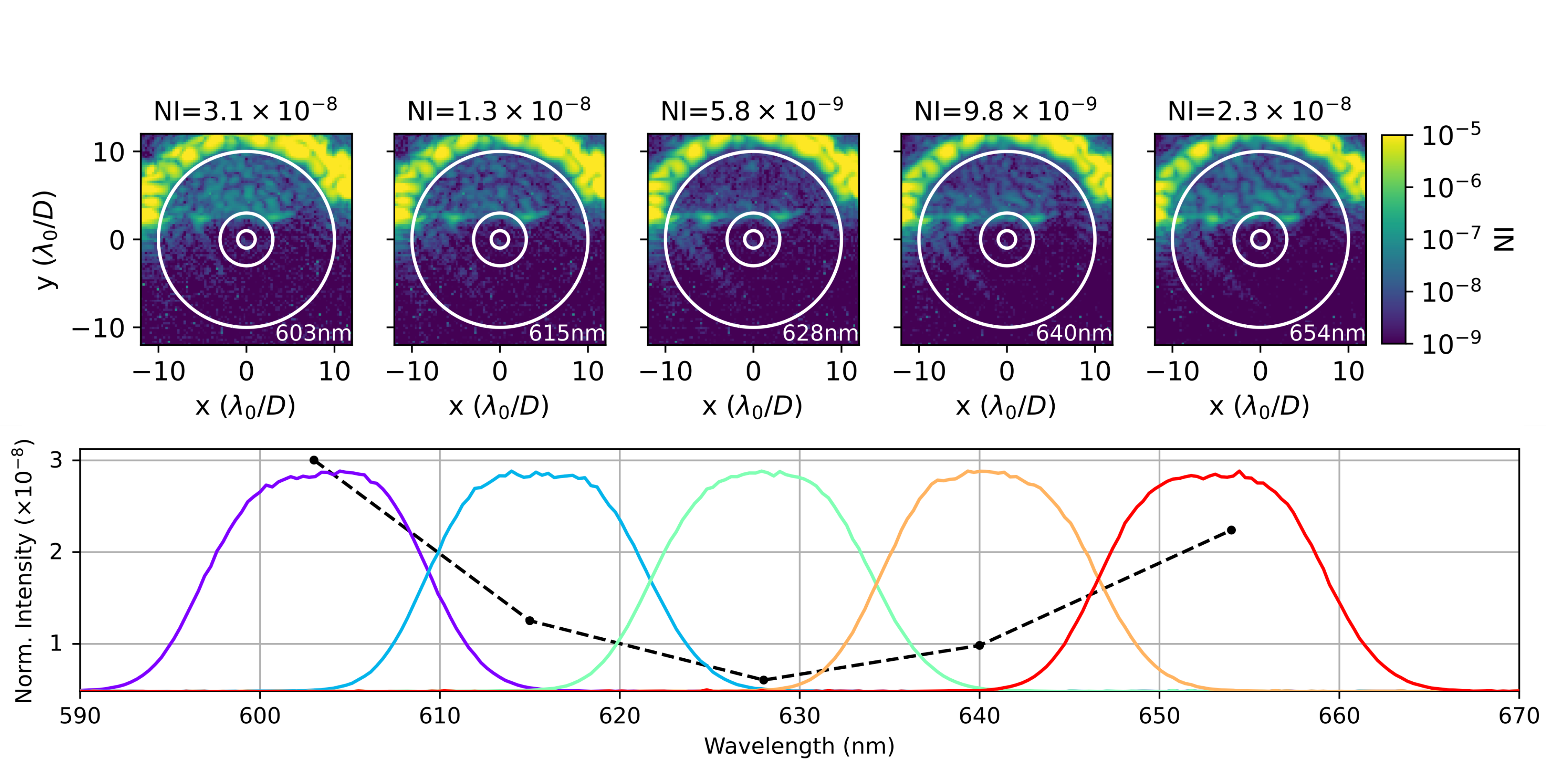}
    \caption 
    { \label{fig:dh_vacuum_bands} Top: NI images of each of the \glspl{DH} in 5 individual 2\% bands over a 10\% bandwidth centered at $\lambda_0$=630nm. Radial separations of 1, 3, and 10 $\lambda_0/D$ are marked. Bottom: The mean \gls{NI} (dashed line) at each center wave, with the measured spectrum for the 2\% bands overplotted.} 

\end{figure}

\begin{figure}[!htb]
    \centering
    \includegraphics[width=\textwidth]{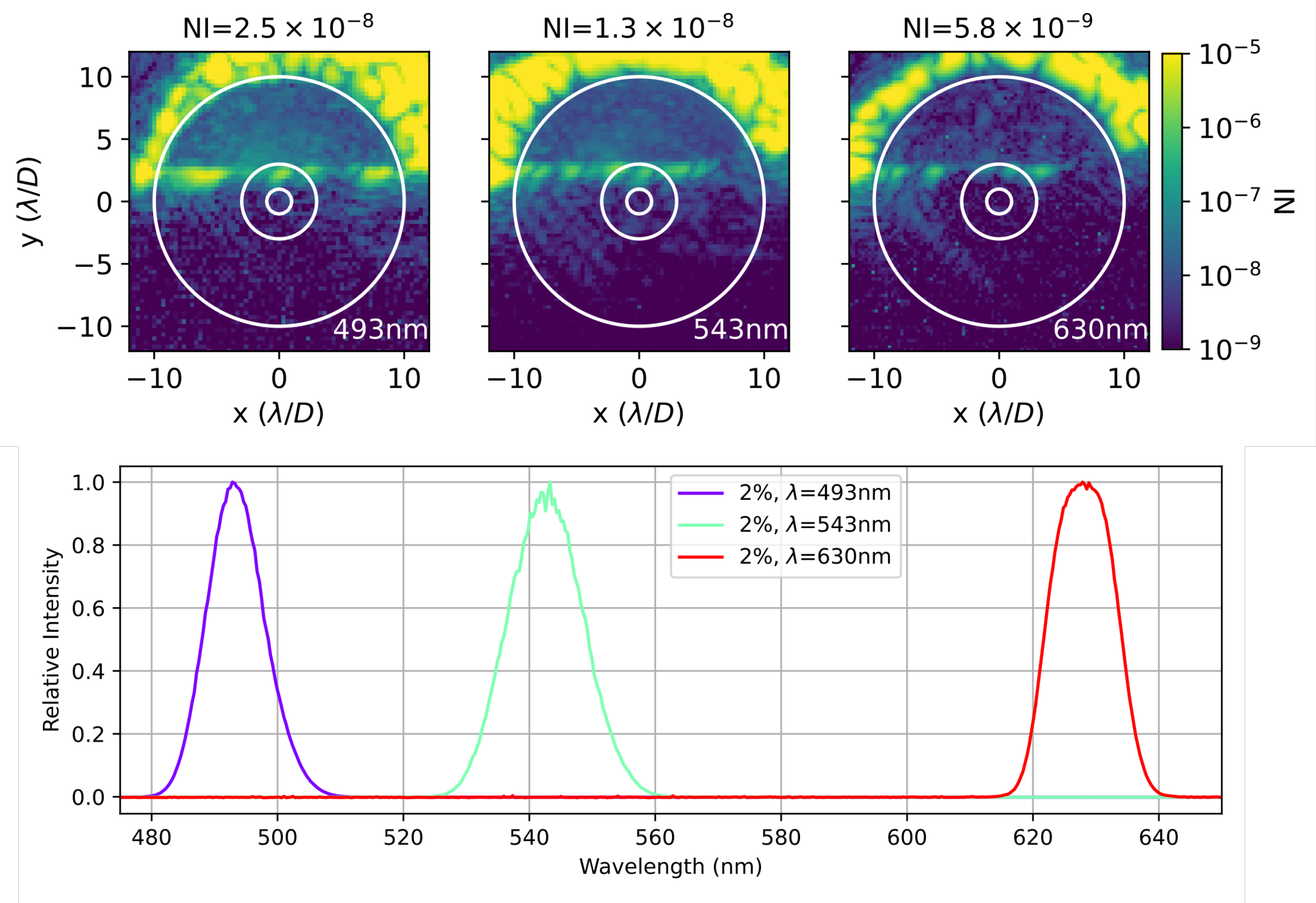}
    \caption{Top: Dark holes for three 2\% bandpasses at central wavelengths of 493, 543, and 630nm. The mean \gls{NI} in a half-sided DH from 3-10$\lambda/D$ is reported. Radial separations of 1, 3, and 10 $\lambda_0/D$ are marked. Bottom: Measured spectra for each of the bandpasses considered here.}
    \label{fig:dh_shortwave}
\end{figure}

A summary of the best contrast performance is given in Table \ref{table:contrast_summary} and in more detail in Figures \ref{fig:dh_air} - \ref{fig:dh_shortwave}. The deepest \gls{NI} achieved to date on the testbed is $2.2\times10^{-9}$ from $3-10 \lambda/D$ in a narrow band ($\ll 1\%$) centered at 630nm. This result was achieved in vacuum, although we have found that similar performance is possible in air with the TVAC chamber sealed.

We explored the spectral dependence of the contrast in a few different ways. In Figure \ref{fig:dh_air}, we performed the iEFC \gls{WFSC} in a range of bandwidths from 2-15\% at 630nm and evaluated the NI after convergence for each case. In comparison, in Figure \ref{fig:dh_vacuum_bband}, we performed the iEFC WFS\&C step in a narrow band and then adjusted the source bandpass to evaluate the NI with increasing bandwidth. The latter experimental setup showed better performance with increasing bandwidth (about 2x at 15\% BW), which suggests that iEFC may perform better in a narrow sensing and control band. Note, however, that the two configurations are not directly comparable: the broadband digging experiment was performed in air in the absence of a field stop, while the latter experiment was performed in vacuum with a field stop. Further experiments are necessary to definitively determine the performance of iEFC with bandwidth. Following Ruane et al.\cite{ruane_broadband_2022}, we also broke down a 10\% bandwidth into individual 2\% sub-bands in Figure \ref{fig:dh_vacuum_bands} to investigate the spectral dependence of the NI in narrow bands that comprise a larger bandwidth.

For comparison, \gls{VVC} tests on the \gls{DST} show a contrast floor of ${\sim}7\times10^{-10}$ in a 2\% BW \cite{ruane_broadband_2022} compared to our best performance of $4\times10^{-9}$ in a 2\% BW, so the overall contrast performance on \gls{scoob} is about 6x worse. The degradation of contrast with spectral bandwidth, however, is very similar between the two testbeds, in spite of the very different sensing and control strategies (\gls{EFC} in multiple sensing bands and control with two DMs on the DST, compared to iEFC sensing and control in a single band with a single DM on SCoOB). The 10:2\% contrast ratio on the DST is ${\sim}2$, and on SCoOB is ${\sim}2.5$. The 15:2\% ratio on the DST is ${\sim}5$, and on SCoOB is ${\sim}4$.

To explore the performance of \gls{scoob} at bluer wavelengths, we conducted a series of DH-digging exercises in 2\% bandpasses at central wavelengths as short as 493nm (see Figure \ref{fig:dh_shortwave}). Note that, although the SCoOB optics were built to enable coronagraphy into the UV, the VVC unit in use for these experiments was specified to a 20\% BW centered at 635nm (570-700nm), so retardance errors at these out-of-band wavelengths are expected to be large (see Figure \ref{fig:vvc_unit}), and therefore create a significant leakage term through the coronagraph. In spite of this, we converged to a NI of $2.5\times10^{-8}$ at 493nm and $1.3\times10^{-8}$ at 543nm.

\begin{figure}[!htb]
    \centering
	\includegraphics[width=\linewidth]{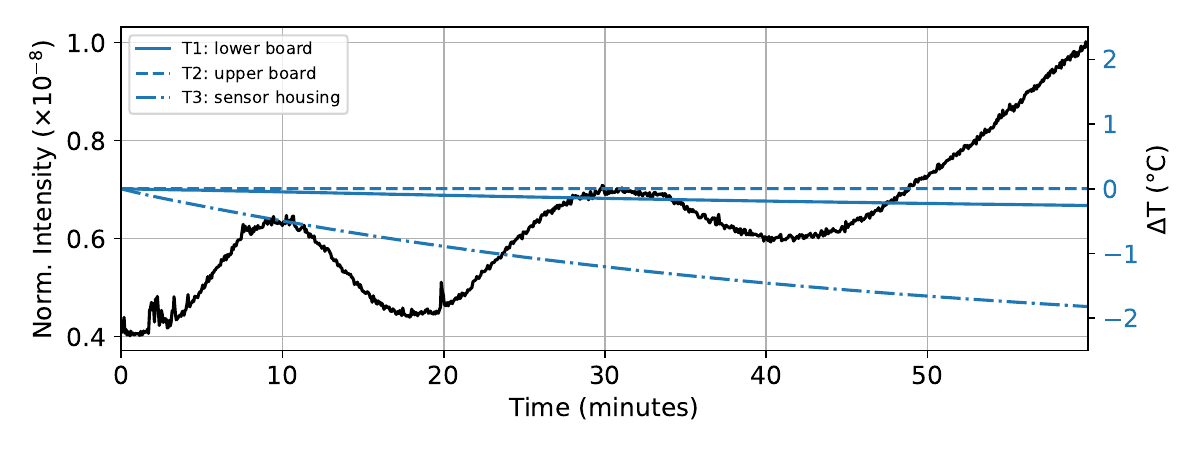}
\caption 
{ \label{fig:dh_stability} \gls{NI} contrast over the course of an hour after digging a \gls{DH} with the narrow-band source. Temperatures at a three locations on the testbed are overplotted.}
\end{figure}

To investigate the stability of the system in vacuum, we dug a \gls{DH} with the narrow-band source to ${\sim}4\times10^{-9}$ \gls{NI} and then monitored the DH over the course of an hour in the absence of any control. In the experiment shown in Figure \ref{fig:dh_stability}, the contrast degrades by a factor of 2.5 over an hour. We also monitor the temperature of the testbed at a handful of locations. The largest temperature swings are observed on the camera housing, which currently has only passive thermal management. As a result, the sensor runs very warm ($>40^{\circ}$C) in vacuum, and the temperature shows large swings as camera settings are changed. We speculate that the sensor creates large and variable thermal gradients on the testbed and is responsible for the majority of the observed contrast instability. Planned mitigations to improve our vacuum stability are outlined in Section \ref{sec:limits}.

\section{Contrast limitations and mitigations}\label{sec:limits}

A number of terms that have limited the testbed contrast at various stages of its development have been identified and---in most cases---mitigated.

\begin{figure}[!htb]
    \centering
    \includegraphics[width=\textwidth]{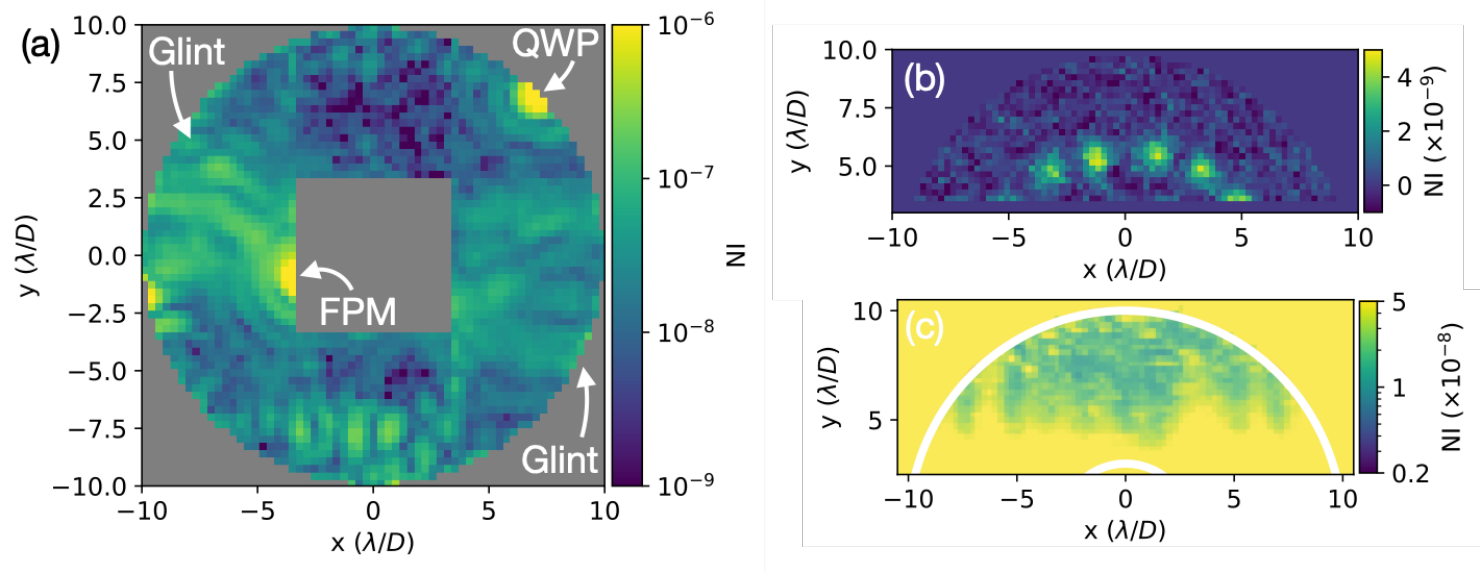}
    \caption{(a) An ensemble of terms that were found to limit the testbed contrast performance in an earlier configuration. The 360$^\circ$ \gls{DH} shown here was constructed by stacking 4 half-sided DHs dug separately in a 10nm-bandwidth centered at 630nm. (b) Speckle pattern that arose from interference on the DM cables with the testbed in vacuum. (c) Pixel cross-talk on the Sony IMX571 from bright speckles outside the DH bleeding into the DH.}
    \label{fig:contrast_limiters}
\end{figure}

Three significant terms that limited broadband performance are shown in Figure \ref{fig:contrast_limiters}(a). In a previous configuration of the testbed, the input linear polarizer and \gls{QWP} were in the collimated space after the \gls{DM}. The QWP created an incoherent ghost at the $10^{-6}$ contrast around 9$\lambda/D$ that would not shift in separation as the QWP was tilted, indicating that it likely arose from an internal back-reflection in the substrates that sandwich the QWP polymer stack. We are working with vendors to fabricate QWPs between wedged substrates that should shift this ghost outside the coronagraph \gls{FOV}. In the meantime, we have mitigated this ghost by placing the input linear polarizer and QWP in the source optics prior to the pinhole, which filters out the ghost.

A second ghost at a similar contrast level was identified around 3$\lambda/D$ and traced to the \gls{VVC}. We mitigated this limitation by swapping to a VVC that showed no evidence of this ghost. A third term identified was a diffuse horizontal structure at the $10^{-7}$ contrast level. This term was traced to a glint from one of the DM actuator print-through speckles off an OAP mount in a defocused plane immediately upstream of the \gls{FPM}. A small re-alignment of the testbed to move the beam away from the offending mount eliminated this structure in the DH.

A speckle pattern showing a structured radial symmetry that arose only intermittently in vacuum is shown in Figure \ref{fig:contrast_limiters}(b). This pattern was traced to two sources of interference with the DM cables---in one case a power cable that was in contact with the cables, and in a second case the cables lying in contact with the chassis of the \gls{COTS} DM electronics box. We speculate that this interference created voltage noise that excited a vibration mode on the DM membrane in vacuum. The speckles appear at ${\sim}5\lambda/D$.

Another intermittent term appeared in both air and in vacuum and was traced to a back reflection from the precision pinhole that coupled back into the fiber and created instability in the frequency-stabilized narrow-line source. This was mitigated by the addition of an inline fiber isolator (see Figure \ref{fig:testbed_layout}.)

An immediate degradation of the contrast floor was discovered in the transition from the Sony IMX492 sensor used for in-air tests to the IMX571 intended for vacuum testing. The worsened contrast was initially attributed to a ghost from the sensor window, but the effect persisted after removal of the window. The field stop (or even a knife edge along placed along the \gls{IWA} of the DH) does reduce or eliminate the effect, and we speculate that it arises from a form of pixel cross-talk that causes the signal from the bright speckles outside the DH to leak into the DH (see Figure \ref{fig:contrast_limiters}(c)). Further investigations are underway to characterize this.

The vacuum performance in the results reported here was limited by drift that presented at the minutes- to hours-timescales, likely due to large thermal gradients created by our passive thermal-strapping solution for the vacuum sensor (see Figure \ref{fig:dh_stability}). Two pending upgrades to the testbed should mitigate this: (1) the integration of closed-loop \gls{LLOWFS} and the addition of a \gls{FSM} for \gls{DH} stabilization, and (2) active cooling of the sensor with a cold plate combined with a mounting solution that thermally isolates the sensor from the bench.

\begin{figure}
    \centering
    \includegraphics[width=\linewidth]{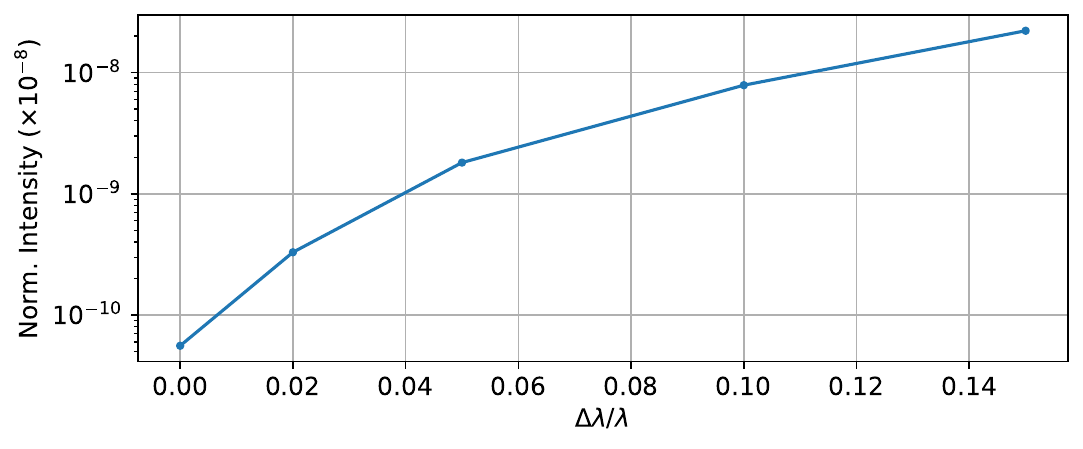}
    \caption{Simulated \gls{NI} contrast against bandwidth for an actuator pinned at a 150nm deflection upstream of a \gls{VVC}. \gls{EFC} compensates for the pinned actuator in monochromatic light, but the correction degrades rapidly with increasing bandwidth.}
    \label{fig:pinned_act}
\end{figure}

A contrast limitation that shows a strong spectral dependence is the presence of pinned actuators on the DM. Our Kilo-C DM has a single actuator pinned at the 0V position. We reduce the effect of this actuator by operating our DM at a low voltage bias (40\%), but the pinned actuator produces a deflection estimated to be ${>}100$nm surface. We simulate an optical system in \texttt{hcipy}\cite{por_hcipy} with an actuator pinned at a 150nm surface deflection upstream of a vortex coronagraph. The pinned actuator creates a ${\sim}10^{-5}$ contrast floor at 630nm, which we compensate for by running \gls{EFC} in a monochromatic band until  a one-sided 3-10$\lambda/D$ DH converges to $<10^{-10}$. The compensation for the pinned actuator degrades with increasing bandwidth. At a 10\% bandwidth, the floor is predicted to approach $10^{-8}$ \gls{NI}. See Figure \ref{fig:pinned_act}. This term could easily explain the majority of our observed spectral contrast performance.

Other suspected sources of contrast limitations include scattering off the light-absorbent foil on the first OAP immediately downstream of the pinhole (which is extremely overfilled in order to select a vary small region of the pinhole beam), scattering off the Aeroglaze Z306 that defines the entrance pupil at the second reflective optic, and spatially-localized defects in the \gls{VVC}, particularly toward the \gls{IWA}\cite{ruane_broadband_2022}.

\section{Conclusions}\label{sec:conclusions}

We have reported recent upgrades to the \gls{scoob} testbed, including the integration of a \gls{VVC}, the expansion of the optical design to incorporate a field stop, and the first results in vacuum. The best contrast performance to-date in a half-sided D-shaped dark hole is $2.2\times10^{-9}$ in a $\ll 1 \%$ BW, $4\times10^{-9}$ in a 2\% BW, and $2.5\times10^{-8}$ in a 15\% BW. We have identified a number of sources that we believe limit our current vacuum performance and plan to integrate solutions to these in the coming months, including better thermal management and isolation of the vacuum camera, closed-loop \gls{LLOWFS} with a FSM and the DM, the fabrication of polarization optics between wedged substrates for ghost control, and improved scattered light mitigation.

Ongoing efforts on the testbed are aimed at integrating flight-like subsystems and validating their performance in a vacuum environment. The current Newport tip/tilt stage will be replaced by a PI-S316 tip/tilt actuator driven by custom vacuum-compatible electronics, the \gls{COTS} DM electronics currently in use will be similarly replaced by custom electronics, a prototype of which was previously demonstrated in SCoOB\cite{vangorkom_dm_2023}. The Lyot mask will shortly be replaced with a reflective Lyot stop that includes an offset pinhole for simultaneous LLOWFS and \gls{SCC} experiments. A new round of \glspl{VVC} with improved retardance performance\cite{anche_pol_scoob_2024} were recently procured and are undergoing characterization. In addition, SCoOB is planned as a facility for vacuum testing of black silicon apodizer masks as part of a funded SAT effort (P.I. A.J. Riggs).

\acknowledgments 

The authors thank Garreth Ruane and Ruslan Belikov for their helpful conversations and invaluable insights. Portions of this research were supported by funding from the Technology Research Initiative Fund (TRIF) of the Arizona Board of Regents
and by generous anonymous philanthropic donations to the Steward Observatory of the College of Science at the University of Arizona.

\bibliography{report,esd} 
\bibliographystyle{spiebib} 

\end{document}